\documentclass[journal]{IEEEtran}

\ifCLASSINFOpdf
  \usepackage[pdftex]{graphicx}

\else

\fi

\usepackage{amsmath,amsfonts,amssymb,amsthm}
\usepackage{mathtools}
\usepackage{commath}
\usepackage[sc,osf]{mathpazo}

\let\norm\undefined 
\DeclarePairedDelimiter\norm{\lVert}{\rVert}

\usepackage{algorithmic}

\usepackage{array}
\usepackage{relsize}
\usepackage{url}

\begin{document}

\title{A Novel Wavelet-base Algorithm for Reconstruction of the Time-Domain Impulse Response from Band-limited Scattering Parameters with Applications}

\author{Shantia~Yarahmadian,~\IEEEmembership{Member,~SIAM}, Maryam~Rahmani,~\IEEEmembership{Member,~IEEE,}
         and~Michael~Mazzola,~\IEEEmembership{Member,~IEEE}
\thanks{M. Rahmani is with the Department
of Electrical and Computer Engineering, Indiana University-Purdue University, Indianapolis,
IN, 46202 USA e-mail: (see http://engr.iupui.edu/main/people/detail.php?id=maryrahm).}
\thanks{M. Mazzola is with University of North Carolina, Charlotte as the director of EPIC, NC, 28223 USA e-mail: (see https://epic.uncc.edu/directory/mike-mazzola).}
\thanks{Sh. Yarahmadian is with Department of Mathematics, Mississippi State University, MS, 39762 USA e-mail: (http://math.msstate.edu/research/bio.php?rec=367).}}


\maketitle

\begin{abstract}
In this paper, we introduce a novel wavelet-based algorithm for reconstructing time-domain impulse responses from band-limited scattering parameters (frequency-domain data) with a particular focus on ship hull applications. We establish the algorithm and demonstrate its convergence, as well as its efficiency for a class of functions that can be expanded as exponential functions. We provide simulation results to validate our numerical results.
\end{abstract}


\IEEEpeerreviewmaketitle

\section{Introduction and Motivation: Ship Hull with References}\label{Intro}

\IEEEPARstart{N}{ew} Naval ships differ significantly from commercial ones as they require a larger amount of energy and power to support applications such as pulse weaponry and high-power military loads. This demand for higher installed power necessitates novel energy conversion and power delivery systems. Military applications have specific requirements, including low signatures to avoid detection by enemies, non-interference, and damage tolerance for recovery and sustainability. Regular electric drive systems comprise power generation, distribution, and control sections, while electric ships must also include pulse power and pulse energy weaponry, communication, computer, radar and sonar systems, electromagnetic assistance launch, hospitality, and service loads. Power systems in electric ships should be reliable and operational even after sustaining damage in an attack. Future all-electric ships differ from terrestrial distribution systems in terms of power density, load characteristics, physical dimensions, characteristic impedances, reliability, and availability. Since the ship hull has some conductive structures, various modes of common-mode currents are induced by current-carrying conductors that are in proximity to the ship hull, which is categorized as an ungrounded/high impedance grounded power system. These coupling modes may affect the level of overvoltages and high-frequency ringing, the electromagnetic signature of the ship, as well as unwanted interference with cathodic protection or the degaussing system \cite{ESRDC_report},\cite{bash2009medium}, \cite{graber2010validation}, \cite{rahmani2015modeling}, \cite{mazzola2015behavioral}, \cite{graber2015scattering}. \par

\begin{figure}[!t]
\centering
\includegraphics[width=1.8in]{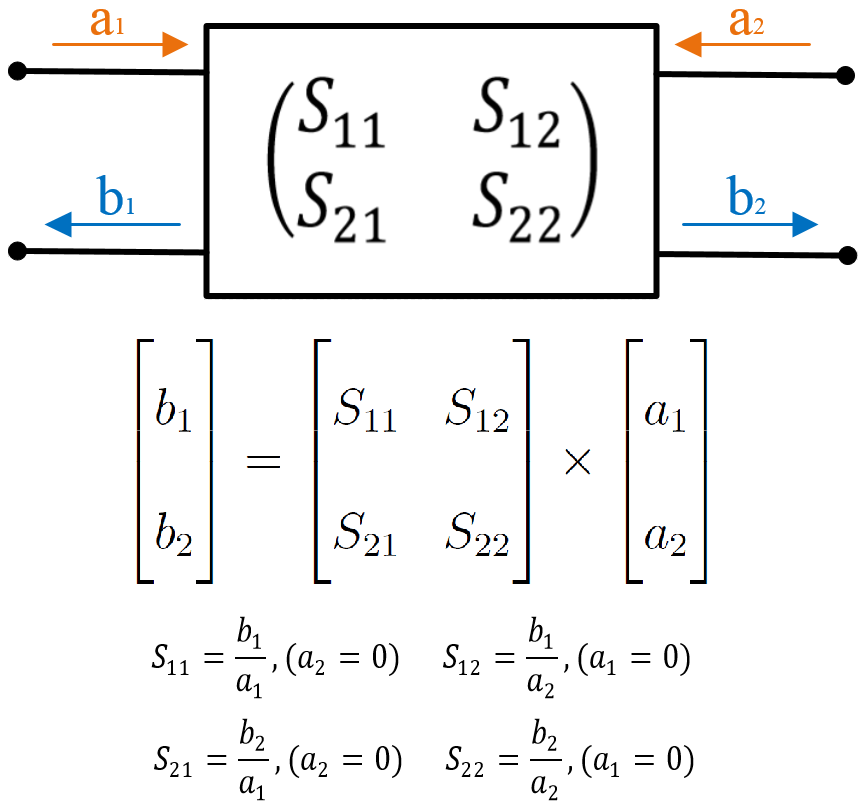}
 \caption{Illustration a two-port network and the related S-parameter matrix.}
\label{Spara}
\end{figure}

The goal of this study is to establish a mathematical framework for coupling transient simulation modes of the shipboard power system with finite element models of the ship hull. In 2010, ESRDC's grounding team introduced Scattering parameters (S-parameters) to transient power system models as a means of evaluating the suitability of component models for grounding studies \cite{graber2010validation}. This method can calculate all conductive, inductive, and capacitive couplings explicitly. S-parameters are network parameters in the frequency domain that describe the transmission and reflection of incident waves at each port of an n-port linear time-invariant network for a given steady-state stimulus. S-parameters represent the response of an N-port network to a voltage signal at each port, where $S_{ij}$ denotes the incident port j and the responding port i. For example, $S_{21}$ represents the incident voltage applied to port 1, and the responding port is 2. The configuration of the S-parameter matrix can be illustrated in Figure \ref{Spara}, where the voltage at the incident port is denoted by 'a' and the voltage at the responding port is denoted by 'b'.

Later, S-parameters were utilized to model common-mode coupling of components of the shipboard power system to the ship hull \cite{rahmani2015modeling}. S-parameters of a power system can be calculated using numerical methods such as finite element analysis (FEA) or by measurement utilizing network analyzers. However, due to the enormous size of the ship hull, it is not possible to measure its S-parameters directly. Thus, a smaller model is employed to validate the capability of S-parameters in determining the transient behavior of the system. For instance, in \cite{mazzola2015behavioral}, the behavioral modeling of a high-power multi-chip module (MCM) package layout and die integration is calculated using S-parameters, and the spectrum measurement by network analyzer confirms the simulated results, showing the reliability of S-parameter models.

It is worth noting that S-parameters are a function of frequency, and band limitation is a common issue in their measurement or simulation. The lower frequency limits for the simulated and measured S-parameters are constrained due to the limitations of simulation algorithms and measuring equipment, leading to missing simulated or measured data, particularly the Dm and close to DC components. This results in a band-limited frequency data set. In ship hull applications, the capacitive nature of coupling to ground results in the lower frequency limit being as high as in the kilohertz range. The lower limit of the frequency range can be determined by the noise floor level of the network analyzer, as mentioned in the manual, and the characteristics of the network under test. In simulation, the lower limit of the frequency band is determined by the robustness of the numerical algorithm.

To obtain an accurate impulse response in the time domain, a complete dataset of S-parameters can be reconstructed by applying the inverse Fourier transform. However, the existing frequency dataset is band-limited, and applying the inverse Fourier transform can lead to certain issues that require special consideration and robust and new reconstruction methods, as studied in the following sections.

In summary, the paper is organized as follows: Section 1 provides the introduction; Section 2 describes the methodology used; Section 3 addresses the limitations of the current approach; Section 4 presents and discusses the proposed reconstruction methods; and finally, Section 5 concludes the paper.

\section{Reconstruction in the Time-domain: A short summary of the previous methods}\label{methods}
Reconstruction of the time domain signal from its frequency domain data (S parameter) can be achieved via inverse Fourier transform:
\begin{equation}\label{eq:1}
{\small
\begin{aligned}
f(t) = \frac{1}{{2\pi }}\int_{ - \infty }^{ + \infty } F\left( \omega  \right){e^{j\omega t}}d\omega
\end{aligned}}
\end{equation}

When reconstructing the time-domain signal from its frequency domain data (S parameter) using inverse Fourier transform, the result is not accurate for band-limited frequency data sets. This is because the DC and close to DC components are missing, which can lead to issues with causality, passivity, and time delay. Specifically, the band-limited nature of the S-parameter model violates the Kramers-Kronig relations, resulting in a non-causal time-domain response. While techniques such as hypothetical causality correction and windowing can be used to address this issue, they often have limited effectiveness in improving the accuracy of time-domain simulations.

\subsection{Causality:}
Causal systems satisfy Kramers-Kronig (K-K) equations:
{small
\begin{equation}\label{eq:2}
u(\omega)=\frac{1}{\pi} P \int_{-\infty}^{+\infty} \frac{v(\omega \prime)}{\omega -\omega \prime} d \omega \prime
\end{equation}}
{\small
\begin{equation}\label{eq:3}
v(\omega)=\frac{1}{\pi} P \int_{-\infty}^{+\infty} \frac{u(\omega \prime)}{\omega -\omega \prime} d \omega \prime
\end{equation}}

The S-parameter model consists of the real part $u(\omega)$ and the imaginary part $v(\omega)$, with P representing the Cauchy principal value. However, because the S-parameter model is band-limited, it violates the Kramers-Kronig (K-K) relations, resulting in a non-causal time-domain response. Although the hypothetical causality correction method is one way to correct the non-causal $(t<0)$ part of the response by truncating it to zero, the resulting spectrum of the truncated response does not match the original one, leading to poor time-domain simulation accuracy \cite{rao2008need, rao2011optimization}. The windowing technique is another way to enhance a signal with ripples, but it can have adverse effects on causality since popular windows like Hanning, Hamming, and Bartlett are non-causal. Despite smoothing out the ripples, applying these windowing techniques can result in a significant portion of signal energy appearing at $t<0$. Biernacki et al. \cite{biernacki2009causality} address causality enforcement in fast EM-based simulation of multilayer transmission lines.

\subsection{Passivity:}
Reconstructing a signal in the time-domain from its Fourier transform can cause passive component models to appear non-passive due to errors in measurement or simulation. However, in the S-parameter matrix, if the magnitude of all the eigenvalues is less than one, it can be concluded that the matrix belongs to a passive component \cite{rao2008need}. One simple way to enforce passivity is to divide the matrix elements at the offending frequencies by the magnitude of the maximum matrix eigenvalue \cite{rao2011optimization}.

\subsection{Time-Delay:}
Time-delay is a response that takes into account the minimum propagation delay. However, this constraint is often violated in numerical simulations due to frequency-domain data truncation or band-limiting, which results in the violation of the Kramers-Kronig criteria.

\section{Previous Methods of Reconstruction}\label{previous}
The previous methods of reconstruction can be categorized into two main methods: reconstruction by circuit elements and reconstruction by curve fitting. In reconstruction by circuit elements method, passive component models are derived using a quasi-static $2D$ and $3D$ solver to find the equivalent inductance $(L)$, mutual inductance $(K)$, and capacitance $(C)$, which are then represented as a transmission line model. The sub-circuit models can be run in transient simulation along with other active device models and are inherently passive and causal. However, these models have some disadvantages, such as incomplete capturing of frequency-dependent loss and dispersion, inadequate modeling of longitudinal modes and transverse currents, and an increase in model run-time and complexity with shorter cross-section lengths.
\\
One of the traditional simulation tools used to solve transient simulations through time stepping differential equations is the generic SPICE. To use this simulator, S-parameters must be converted to a lumped element circuit model using algorithms such as Genetic Algorithm, Macro-Modeling, and Laplace domain simulation. Laplace simulation uses recursive convolution to evaluate the model in one of the following forms: Pole/Residue, Pole/Zero, or Rational Polynomial.
\\
In the reconstruction using curve fitting, missing parts of S-parameter data can be extracted and the time-domain impulse response can be reconstructed through simple IDFT. Popular curve fitting software and toolboxes can be found in Excel, Matlab, and Keysight (Agilent) ADS. Matlab is a preferred choice because of its exceptional features and its strength in fitting data, curves, and surfaces interactively. The library model types for curves and surfaces include Weibull distribution, exponential functions, Fourier series, Gaussian distribution, interpolants, polynomials, power series, rational equations, sum of sine functions, and splines.
\\
However, the problem with these methods is that they fit a circuit-based model, a curve, a polynomial, or a rational function to the S-parameter data, and they are not able to find a unique curve for frequency data. To estimate a more precise curve, the number of pole-zeros, order of polynomial, and number of elements in the circuit model should be increased, which makes the numerical computations more complex and time-consuming.

\section{Wavelet Based Reconstruction Algorithm}\label{algorithm}
\subsection{Description of the Problem:}
Wavelet-based signal processing has been a well-known and widely used technique for decades \cite{daubechies1992ten}. Typically, the wavelet transform is applied to images in the x-y plane and to time-domain signals that are real-valued \cite{daubechies1992ten}. However, the objective of this work is to reconstruct a time-domain signal using frequency-domain data. To achieve this, the wavelet transform is applied to frequency-domain signals that are limited in bandwidth and have complex values.

\subsection{Band-limited versus time-limited:}
\begin{figure}[!t]
\centering
\includegraphics[width=1.6in]{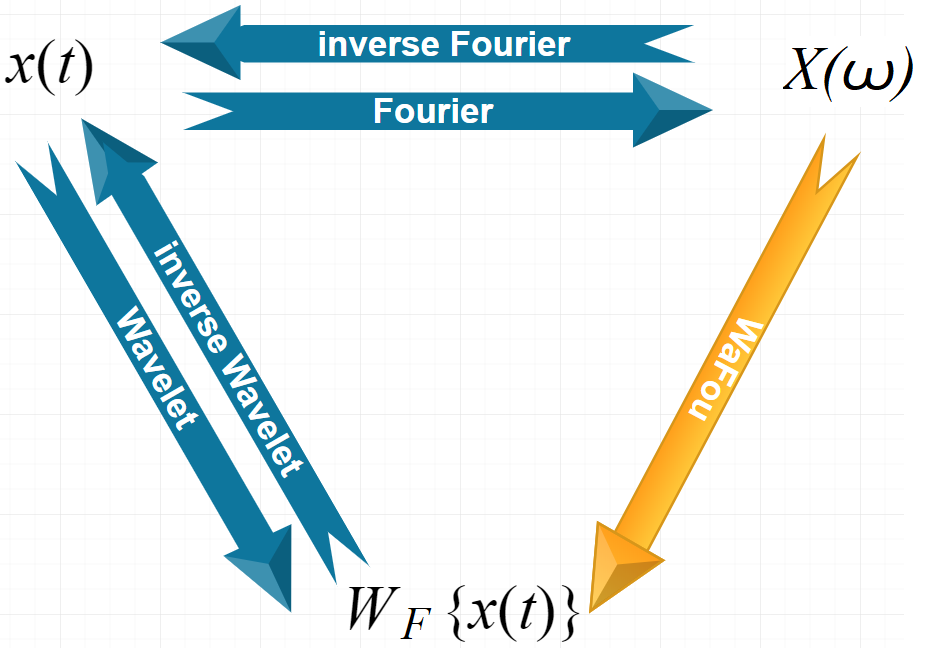}\hfill
 \caption{The schematic diagram of the proposed method}
\label{Loop}
\end{figure}

As mentioned earlier, the Fourier transform (iFFT) can accurately reconstruct a signal in the time domain, provided that the signal is not band-limited. However, since frequency measurement equipment such as network analyzers and frequency-domain simulation software have limitations, the signals are typically band-limited, and applying the iFFT may not yield accurate results. Therefore, there is a need to develop a robust and reliable method for more accurate time-domain signal reconstruction. The proposed method is illustrated in the schematic diagram shown in Fig. \ref{Loop}.

To implement this algorithm, we first establish a mathematical relationship to determine the continuous wavelet transform of the signal from its frequency-domain data. We then apply the continuous wavelet transform to the frequency data to reconstruct the wavelet transform of the signal. Next, we apply an inverse wavelet transform to the resulting signal to reconstruct it in the time domain. After applying the Fourier transform to the reconstructed signal, we compare the resulting spectrum to the original signal's spectrum and adjust the components other than the DC component to their original values. We then multiply the level of difference between the two closest components after the DC component in the original and reconstructed signals by a coefficient, and consider it as the DC component. We repeat this process until the length of the mother wavelet is equal to the length of the signal. The algorithm for this method is explained in the following section.

\section{Mathematical Background}\label{Mathematics}
Signals can be classified as either band-limited (frequency-limited) or time-limited, but not both. When $f(t)$ is a band-limited signal with a finite bandwidth, it is the restriction to $\mathbb{R}$ of an analytic function. If $f(t)$ is also time-limited, with its support in $[-T,T]$ where $T<\infty$, then any nontrivial analytic function can only have isolated zeros. Hence, it is concluded that $f(t)\equiv 0$. Nonetheless, in many applications, there exists an effective scenario of both band-limiting and time-limiting \cite{daubechies1992ten}.

\subsection{Continuous Wavelet and Wavelet Inverse Transform} \label{cwt}
The continuous wavelet transform in the time-domain is defined as:
{\small
\begin{equation}\label{eq:4}
\begin{aligned}
{\cal W}\left\{ {f\left( t \right)} \right\} = {{\cal W}_\psi }\left\{ {f\left( t \right)} \right\} = \int_{ - \infty }^{ + \infty } f\left( t \right)\overline {{\psi _{a,b}}\left( t \right)} dt
\end{aligned}
\end{equation}
where
{\small
\begin{equation}\label{eq:5}
\begin{aligned}
{\psi _{a,b}}\left( t \right) = \frac{1}{{\sqrt {\left| a \right|} }}\psi \left( {\frac{{t - b}}{a}} \right)
\end{aligned}
\end{equation}}

is the mother wavelet. The inverse wavelet transform is defined by the resolution of identities formula, as follows:
{\small
\begin{equation}\label{eq:6}
\begin{aligned}
f\left( t \right) = {{\cal W}^{ - 1}}\left\{ {{\cal W}f\left( t \right)} \right\} = & \\ \frac{1}{{{C_\psi }}}\;\int_{ - \infty }^{ + \infty } \int_{ - \infty }^{ + \infty } f\left( t \right),{\psi ^{a,b}}\left( t \right){\psi ^{a,b}}\left( t \right)\frac{{da\;db}}{{{a^2}}} = & \\ \frac{1}{{{C_\psi }}}\;\int_{ - \infty }^{ + \infty } \int_{ - \infty }^{ + \infty } {{\cal W}_\psi }\left( {a,b} \right)\frac{1}{{\sqrt a }}\psi \left( {\frac{{t - b}}{a}} \right)\frac{{da\;db}}{{{a^2}}}\;\;\;
\end{aligned}
\end{equation}}
where
{\small
\begin{equation}\label{eq:7}
\begin{aligned}
{C_\psi } = \;\int_{ - \infty }^{ + \infty } \frac{{{{\left| {\Psi \left( \omega  \right)} \right|}^2}}}{{\left| \omega  \right|}}d\omega  < \infty
\end{aligned}
\end{equation}}

Note that, $\Psi(\omega)$ is the Fourier transform of the mother wavelet $\psi_{a,b} (t)$ and  $,$ stands for the $L_2$- inner product. $C_\psi$ is the admissible constant and should satisfy $0<C_\psi<\infty$, in other words it should be a function with compact support.

\subsection{From Fourier Transform to Wavelet Transform} \label{wafou}
Parsvale’s identity states that the total energy of the signal in the time-domain is equal to the energy in the frequency-domain up to $\frac{1}{2\pi}$. Therefore:
{\small
\begin{equation}\label{eq:8}
\begin{aligned}
\int_{ - \infty }^{ + \infty } {f_1}\left( t \right)\overline {{f_2}\left( t \right)} \;dt = \frac{1}{{2\pi }}\int_{ - \infty }^{ + \infty } {F_1}\left( \omega  \right)\overline {{F_2}\left( \omega  \right)} \;d\omega
\end{aligned}
\end{equation}}
If f(t) is square integrable on $(-\infty,+\infty)$, and $\psi(t)\in L^2 (\rm I\!R)$ then the continuous wavelet exists \cite{daubechies1992ten}. Using Parsval’s identity it can be concluded:
{\small
\begin{equation}\label{eq:9}
\begin{aligned}
 W\left\{ {f\left( t \right)} \right\} =\;\int_{ - \infty }^{ + \infty } f\left( t \right)\overline {\psi \left( {\frac{{t - b}}{a}} \right)} \;dt =\\
 \frac{1}{{2\pi }}\;\int_{ - \infty }^{ + \infty } F\left( \omega  \right)\sqrt {a}     \overline {{e^{ - ib\omega }}} \;\;\overline {{\rm{\Psi }}\left( {a\omega } \right)} \;d\omega \left( {{W_\psi }f} \right)\left( {a,b} \right) = \\
 \;\frac{1}{{2\pi }}\;\int_{ - \infty }^{ + \infty } F\left( \omega  \right)\sqrt {a} {e^{ + ib\omega }}\;\;\overline {{\rm{\Psi }}\left( {a\omega } \right)} \;d\omega
\end{aligned}
\end{equation}}

Where $\Psi(\omega)$ stands for the Fourier transform of $\psi(t)$.

\textbf{Theorem 1}
For all $f,g,\psi(t)\in L^2 (\rm I\!R)$
{\small
\begin{equation}\label{eq:10}
\begin{aligned}
\int_{ - \infty }^{ + \infty } \int_{ - \infty }^{ + \infty } \left( {{W_\psi }f} \right)\left( {a,b} \right)\overline {\left( {{W_\psi }g} \right)}\left( {a,b} \right) \frac{{da\;db}}{{{a^2}}} = {C_\psi }\langle f,g\rangle\;\;\;
\end{aligned}
\end{equation}}
where
{\small
\begin{equation}\label{eq:11}
\begin{aligned}
{C_\psi } = 2\pi \int \frac{{{{\left| {\hat \psi \;\left( \xi  \right)} \right|}^2}}}{{\left| \xi  \right|}}d\xi  < \infty
\end{aligned}
\end{equation}}
Proof: See \cite{daubechies1992ten}.

\subsection{Reconstruction} \label{reconstr}
By applying formula \eqref{eq:9} and \eqref{eq:10}, we develop the method for reconstructing time signal from the frequency data, via wavelet, as follows:

\textbf{Theorem}

\begin{equation}\label{eq:12}
{\tiny
\begin{aligned}
\lim_{\substack{A_1\to\infty \\ A_2,B\to\infty \\ \Omega_1\to+\infty , \Omega_2\to-\infty}} \norm[\big]{f(t)-\frac{1}{C_\psi}\iint\limits_{\substack{A_1\leq |a|\leq A_2 \\ |b|\leq B}}\Big(\frac{1}{2\pi}\int_{\Omega_1}^{\Omega_2}F(\omega) \sqrt{|a|} \times
e^{+ib \omega} \overline { {\Psi(a \omega) }}\left( {a,b} \right) \frac{{da\;db}}{{{a^2}}} \Big)}
\end{aligned}}
\end{equation}

\textbf{Proof}
The absolute value of the inner product is bounded. Therefore:
{\begin{equation}\label{eq:13}
{\small
\begin{aligned}
\iint\limits_{\substack{A_1\leq |a|\leq A_2 \\ |b|\leq B}}\frac{{da\;db}}{{{a^2}}} (\norm{f} \norm{g} \norm{ \psi^{a,b}}= 4B \Big( \frac{1}{A_1} - \frac{1}{A_2}\Big) \norm{f} \norm{g}
\end{aligned}}
\end{equation}

By applying Riesz’ lemma \cite{daubechies1992ten}, we have:

\begin{equation}\label{eq:14}
{\tiny
\begin{aligned}
\norm[\big]{f(t)-\frac{1}{C_\psi}\iint\limits_{\substack{A_1\leq |a|\leq A_2 \\ |b|\leq B}}\Big(\frac{1}{2\pi}\int_{\Omega_1}^{\Omega_2}F(\omega) \sqrt{|a|} \times
e^{+ib \omega} \overline { {\Psi(a \omega) }} d\omega \left( {a,b} \right) \psi^{a,b} \frac{{da\;db}}{{{a^2}}} \Big)}=
\end{aligned}}
\end{equation}

\begin{equation}\label{eq:15}
{\tiny
\begin{aligned}
\sup_{\norm{g}=1} \norm[\big]{\big \langle f(t)-\frac{1}{C_\psi}\iint\limits_{\substack{A_1\leq |a|\leq A_2 \\ |b|\leq B}}\big( \frac{1}{2\pi}\int_{\Omega_1}^{\Omega_2}F(\omega) \sqrt{|a|} \times
e^{+ib \omega} \overline { {\Psi(a \omega) }} d\omega \big) \left( {a,b} \right)\psi^{a,b} \frac{{da\;db}}{{{a^2}}},g \big \rangle} \leq \\ \sup_{\norm{g}=1} \norm[\big]{\frac{1}{C_\psi}\iint\limits_{\substack{A_1\leq |a|\leq A_2 \\ |b|\leq B}} \big(\frac{1}{2\pi}\int_{\Omega_1}^{\Omega_2}F(\omega) \sqrt{|a|} \times e^{+ib \omega} \overline { {\Psi(a \omega) }} d\omega \big) \left( {a,b} \right)\overline{(W_\psi g)(a,b)} \frac{{da\;db}}{{{a^2}}}}\\
\leq \sup_{\norm{g}=1} \norm[\big]{\frac{1}{C_\psi}\iint\limits_{\substack{A_1\leq |a|\leq A_2 \\ |b|\leq B}}  \bigg| \big(\frac{1}{2\pi}\int_{\Omega_1}^{\Omega_2}F(\omega) \sqrt{|a|} \times e^{+ib \omega} \overline { {\Psi(a \omega) }} d\omega \big) (a,b)  \bigg|^2 \frac{{da\;db}}{{{a^2}}}}^\frac{1}{2} \times \\
\norm[\big]{\frac{1}{C_\psi}\iint\limits_{\substack{A_1\leq |a|\leq A_2 \\ |b|\leq B}} \overline{|(W_\psi g)(a,b)|^2} \frac{{da\;db}}{{{a^2}}}}^\frac{1}{2}
\end{aligned}}
\end{equation}

Now by using Theorem 1 (setting $f=g$) and the fact that $\norm{g}^2=1$, the second term equals $1$. Therefore:

\begin{equation}\label{eq:16}
{\tiny
\begin{aligned}
\norm[\big]{f(t)-\frac{1}{C_\psi}\iint\limits_{\substack{A_1\leq |a|\leq A_2 \\ |b|\leq B}}\big( \frac{1}{2\pi}\int_{\Omega_1}^{\Omega_2}F(\omega) \sqrt{|a|} \times e^{+ib \omega} \overline { {\Psi(a \omega) }} d\omega \big) (a,b) \psi^{a,b} \frac{{da\;db}}{{{a^2}}}} \\
\leq \norm[\big]{\frac{1}{C_\psi}\iint\limits_{\substack{A_1\leq |a|\leq A_2 \\ |b|\leq B}} \bigg| \big(\frac{1}{2\pi}\int_{\Omega_1}^{\Omega_2}F(\omega) \sqrt{|a|} \times e^{+ib \omega} \overline { {\Psi(a \omega) }} d\omega \big)(a,b)\bigg|^2 \frac{{da\;db}}{{{a^2}}}}^\frac{1}{2}\\
=\norm[\big]{\frac{1}{C_\psi}\iint\limits_{\substack{A_1\leq |a|\leq A_2 \\ |b|\leq B}} | (W_\psi f)(a,b)|^2 \frac{{da\;db}}{{{a^2}}}}^\frac{1}{2}
\end{aligned}}
\end{equation}

\textbf{Remark 1}
It should be noted that:

\begin{equation}\label{eq:17}
\begin{aligned}
| (W_\psi f)(a,b)|\leq \norm{f}   \quad   , \quad \norm{\psi(t)}=1
\end{aligned}
\end{equation}

\textbf{Remark 2}
The connection between Fourier transform of the signal and its wavelet transform is established through the following formula.

{\small
\begin{equation}\label{eq:18}
\begin{aligned}
{W_\psi }f = \frac{1}{{2\pi }}\;\int_{ - \infty }^{ + \infty } F\left( \omega  \right)\sqrt {\left| a \right|} {e^{ + ib\omega }}\;\;\overline {{\rm{\Psi }}\left( {a\omega } \right)} \;d\omega
\end{aligned}
\end{equation}}

\begin{equation}\label{eq:19}
{\tiny
\begin{aligned}
f\left( t \right) = {W^{ - 1}}\{ ({W_\psi }f)\left( {a,b} \right)\} \left( t \right) = \;\frac{1}{{2\pi {C_\psi }}}\int_{ - \infty }^{ + \infty } \int_{ - \infty }^{ + \infty } \frac{1}{{\sqrt {\left| a \right|} }}\psi \left( {\frac{{t - b}}{a}} \right) \times \left( {\int_{ - \infty }^{ + \infty } F\left( \omega  \right){e^{ + ib\omega }}\overline {\Psi \left( {a\omega } \right)} d\omega } \right)\;\frac{{da\;db}}{{{a^2}}}
\end{aligned}}
\end{equation}

\section{Description of the Algorithm}
The process begins with the frequency data, which is obtained from the spectrum of the signal. The frequency data, also known as S-parameters, is band-limited and may be missing the DC component. To address this, the proposed method constructs the missing DC frequency element and calculates the system impulse response. A flowchart illustrating the method is shown in Figure \ref{flowchart}, which involves the following steps:

\begin{figure}[!t]
\centering
\includegraphics[width=3.6in]{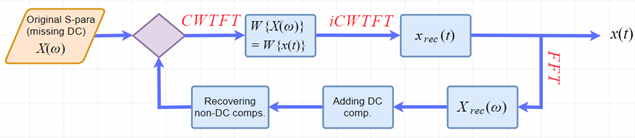}
 \caption{The flowchart of the proposed recursive algorithm.}
\label{flowchart}
\end{figure}

\begin{enumerate}
 \item Extend the real part of the given S-parameter data with even symmetry and the imaginary part with odd symmetry to obtain a real signal in the time domain.
\item Apply the continuous wavelet transform to the signal's spectrum using equation \ref{eq:9}.
\item Apply the inverse continuous wavelet transform to the wavelet coefficients calculated in the previous step to construct the time-domain signal. This is the first iteration, which constructs the missing DC element of the signal.
\item In the next iteration, calculate the spectrum of the reconstructed signal using the Fourier transform. Compare this spectrum to the original signal spectrum and adjust the components other than the DC component to their original values. Multiply the difference between the $k+1$ element in the current and previous spectrum by a scaling factor, where $k$ is the number of missing points (which is one in this example).
\item Apply the CWTFT and iCWTFT to the signal to reconstruct the signal in the time domain.
\item Iterate the above process $M$ times, where $M$ is the length of the scales vector.
  
\end{enumerate}
The reconstructed signal, or the impulse response, is shown in Figure \ref{IR}. The S-parameter matrix elements represent the Fourier transform of the impulse response of the corresponding ports. Therefore, reconstructing the signal in the time domain provides the time-domain impulse response.

\begin{figure}[!t]
\centering
\includegraphics[width=3in]{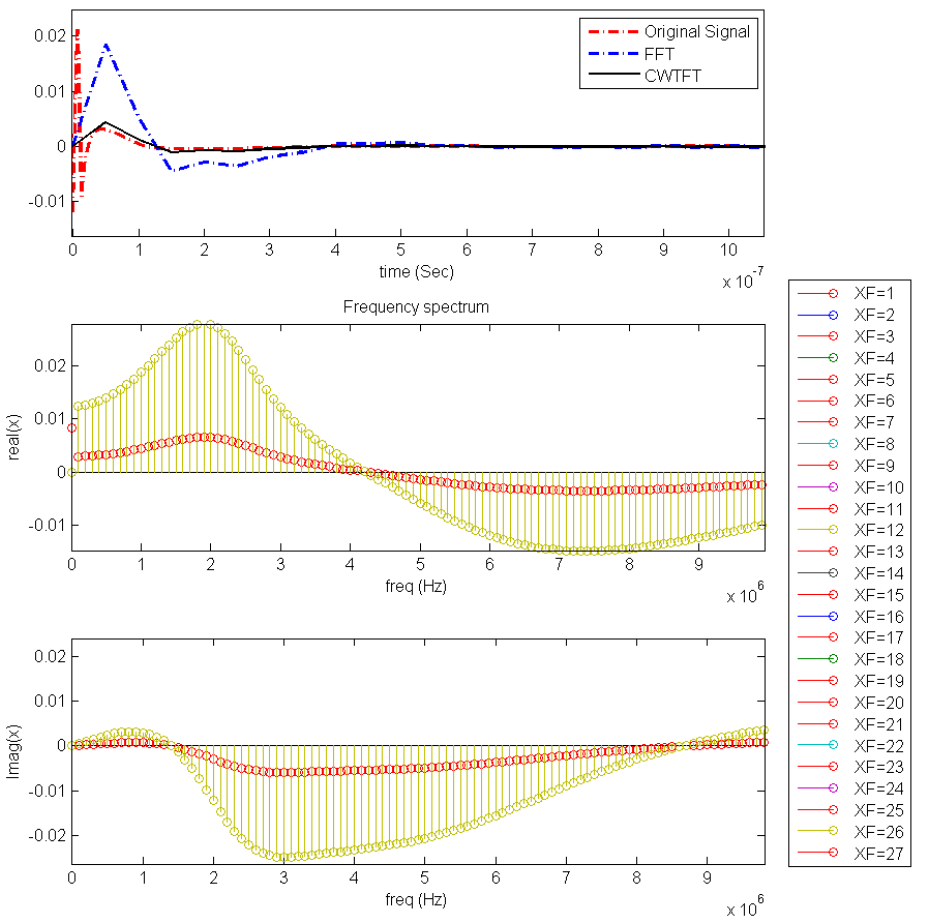}
 \caption{UP: Impulse response, Middle: real part of the spectrum, Down: imaginary part of the spectrum.}
\label{IR}
\end{figure}

\section{Mathematical Representation of the Proposed Algorithm and its Convergency}
We begin with $X(\omega)$, which is the spectrum of the time signal $x(t)$. However, in practice, we start with the S-parameter value $X^0(\omega)$, which is the band-limited version of $X(\omega)$. The goal is to reconstruct the time-domain signal $x(t)$ using the wavelet transform.
\\
In the discrete form, $\omega$ represents the frequency and is band-limited. Therefore, we have $\omega \in {\omega_0, \omega_1, \omega_2, \ldots, \omega_N}$, where $\omega_i \neq 0$ for $i \in {1, 2, \ldots, N}$. We also assume that the missing DC component is $X(\omega_0)$, which is zero at the beginning.

In the first iteration, we find the discrete form of the time-domain signal from the available dataset ${\omega_0, \omega_1, \omega_2, \ldots, \omega_N}$

\begin{equation}\label{eq:20}
\begin{aligned}
\left[ {{{\rm{X}}^0}\left( {{\omega }} \right)} \right] = \left[ {{{{X}}^0}\left( {{{{\omega }}_1}} \right), \ldots ,{{{X}}^0}\left( {{{{\omega }}_{{N}}}} \right)}
 \right]\mathop  \to \limits^{{{First\;Iteration\;}}} \left[ {{{x}}_{{\rm{rec}}}^1\left( {{t}} \right)} \right] = \\ \left[ {{{x}}_{{{rec}}}^1\left( {{{{t}}_1}} \right), \ldots ,{{x}}_{{{rec}}}^1\left( {{{{t}}_{{N}}}} \right)} \right]
\end{aligned}
\end{equation}

Where $X^0(\omega)$ is the given bandlimited frequency-domain spectrum, $X^0(\omega_i)$ is the $ith$ component in the spectrum, $x_{rec}^1 (t)$ is the time-domain reconstructed signal via the first iteration, and $x_{rec}^1 (t_i)$ is the $ith$ component of the  $x_{rec}^1 (t)$.

\textbf{Odd/Even Expansion}
To be able to apply the procedure, we first find expansion of the spectrum $[X^0 (\omega)]$.

\begin{equation}\label{eq:21}
{\small
\begin{aligned}
\left[ {{\rm{\hat X}}\left( {{\omega }} \right)} \right] = \left[ {{{\rm{X}}^0}\left( { - {\omega _{\rm{N}}}} \right), \ldots ,{{\rm{X}}^0}\left( { - {\omega _1}} \right),0,{{\rm{X}}^0}\left( {{\omega _1}} \right), \ldots ,{{\rm{X}}^0}\left( {{\omega _{\rm{N}}}} \right)} \right]
\end{aligned}}
\end{equation}

Where $\hat{X} (\omega)$ is the expansion of  $X^0(\omega)$  such that the real part of $X^0(\omega)$  is expanded evenly and the imaginary part of $X^0(\omega)$  is expanded oddly. Applying the Fourier transform on this expanded signal results in a real signal in the time domain (e.g. $x(t)$) that is the property of every signal in nature.

In the following calculations, we have adopted the following notations, which are described as follows.

$[x_{rec}^1 (t)]=[x_{rec}^1 (t_1 ),…,x_{rec}^1 (t_N )]$ represents the first iteration of the time-domain recovered signal and $[X_{rec}^1 (\omega)]=[X_{rec}^1 (\omega_1 ),…,X_{rec}^1 (\omega_N )]$ denotes its discrete Fourier transform. $ \Delta \omega =|\omega_(i+1)-\omega_i |$, $i\in \{1,2,…,N\}$  is the frequency sampled interval, which is assumed to be uniform. We have considered:

\begin{equation}\label{eq:22}
{\small
\begin{aligned}
{{\rm{s}}_{\rm{n_s}}} = {{\rm{s}}_1}{2^{{\rm{n_s}} \times {\rm{ds}}}} \quad s_1=0.02
\end{aligned}}
\end{equation}

${{\rm{s}}_{\rm{n_s}}}$ is the nth scale value that is calculated  based on the first scale, $s_1$ , and the $ds$ is a constant equal to $0.4875$. The scale in the nth stage show how wide the wavelet function is regarded to the mother wavelet with the width of $s_1$.

\begin{equation}\label{eq:23}
{\small
\begin{aligned}
\left[ S \right] = \left[ {{s_1},{s_2},\; \ldots ,{s_{{n_s}}}} \right] = \left[ {{s_1},{\rm{\;}}{s_1}{2^{ds}}, \ldots ,{s_1}{2^{ {{n_s}} ds}}} \right]
\end{aligned}}
\end{equation}

\begin{equation}\label{eq:24}
{\small
\begin{aligned}
\alpha=\sqrt{\Delta \omega} \times \sqrt{2N-1} \frac{2^m}{\sqrt{m(2m-1)!}}=constant
\end{aligned}}
\end{equation}

$a$ is the Paul wavelet spectrum constant. The time-domain sampling period, $T_s$ is calculated as following:

\begin{equation}\label{eq:25}
{\small
\begin{aligned}
T_s=\frac{2\pi}{\Delta \omega}, \quad  s_1=2T_s, \quad ds=\frac{39}{80}=0.4875, \\
s_12^{n_sds}\leq \frac{N}{\omega_N}\nonumber
\end{aligned}}
\end{equation}
Therefore:
\begin{equation}\label{eq:26}
{\small
\begin{aligned}
n_s \leq \frac{\ln (\frac{N}{s_1 \omega_N})}{ds \ln 2}
\end{aligned}}
\end{equation}

$n_s$ is calculated through the above formula such that it guaranties that the wavelet function is extended in such a scale that covers the entire time-domain signal. For example the length of the time-domain signal can be determined by dividing the number of the frequency components to the maximum of the frequency value, i.e. $t_{max}=\frac{N}{\omega_N}$  , if we consider $N=1000$ and $\omega_N=100$ then the $t_{max}=10$ seconds. Therefore:

\begin{equation}\label{eq:27}
\begin{aligned}
T_s=\frac{t_{max}}{N}=\frac{10}{1000}=0.01, \quad  s_1=2T_s=0.02\nonumber
\end{aligned}
\end{equation}

So plaguing in the values in Eq. \ref{eq:26}, the $n_s =20$ is calculated to cover the entire $10-sec$ time-domain signal as depicted in Fig. \ref{scales}.

\begin{figure}[!t]
\centering
\includegraphics[width=3.5in]{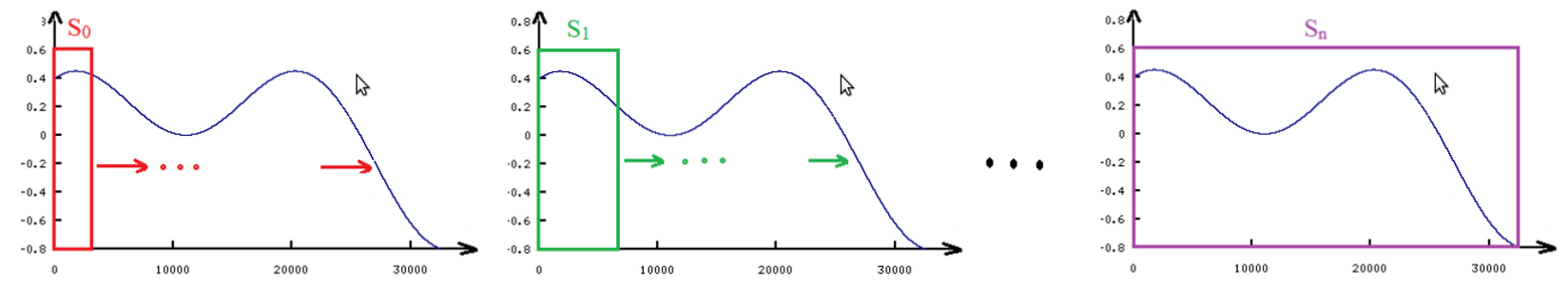}
 \caption{We have the time-domain signal $x(t)$ and a sample mother wavelet. We increase the scales until the mother wavelet covers the entire time-domain signal.}
\label{scales}
\end{figure}

Now, the first iteration of the time domain signal is evaluate by discretizing the continuous form of the Wavelet-Fourier transform introduced in Eq. \ref{eq:19} , which is written as:

\begin{equation}\label{eq:28}
{\small
\begin{aligned}
\left[ {x_{rec}^1\left( t \right)} \right] = \left[ {\begin{array}{*{20}{c}}
{x\left( {{t_0}} \right)}\\
{x\left( {{t_1}} \right)}\\
 \vdots \\
{x\left( {{t_n}} \right)}
\end{array}} \right]=\\ \frac{1}{{2\pi {C_\psi }}} \sum_a \sum_b \frac{1}{{{a^2}\sqrt a }}\left[ {\psi \left( {\frac{{t - b}}{a}} \right)} \right] \times iFFT\left\{ {\left[ {{\rm{\Psi }}_{ij}^1} \right]\left[ {X\left( \omega  \right)} \right]} \right\}
\end{aligned}}
\end{equation}

Note that $iFFT$ is a Matlab built-in function \cite{torrence1998practical}. From now on $a$ is noted as $s$ as the scale in the equations.
where:

\begin{equation}\label{eq:29}
{\small
\begin{aligned}
[\Psi_{ij}^1]=\left[
                \begin{array}{ccccccc}
                  \mu_1(s_1 \omega_1)^m e^{s_1 \omega_1} & \mu_1(s_1 \omega_N)^m e^{s_1 \omega_N} & 0 & \cdots & 0 \\
                  \mu_2(s_2 \omega_1)^m e^{s_1 \omega_1} & \mu_2(s_2 \omega_N)^m e^{s_1 \omega_N} & 0 & \cdots & 0 \\
                  \vdots & \vdots & \vdots & \ddots & \vdots \\
                  \vdots & \vdots & \vdots & \ddots & \vdots \\
                  \vdots & \vdots & \vdots & \ddots & \vdots \\
                  \mu_{n_s}(s_{n_s} \omega_1)^m e^{s_{n_s} \omega_1} & \mu_{n_s}(s_{n_s} \omega_N)^m e^{s_{n_s} \omega_N} & 0 & \cdots & 0 \\
                \end{array}
              \right]\\
              1\leq i \leq n_s, \quad 1 \leq j \leq 2N+1
\end{aligned}}
\end{equation}

 $m$ is the wavelet constant for each mother wavelet and $[s]$  is the scaling vector. Now by applying discrete Fourier transform to $[x_{rec}^1 (t)], [X_{rec}^1 (\omega)]$ can be evaluated. We construct the first iteration of the missing DC part as:
{\small
\begin{equation}\label{eq:30}
\begin{aligned}
X_{rec}^1(\omega_0)=2 \log_{10} (\frac{\Delta \omega}{2 \pi}) \ln (S_1). sign [Re{X^1(\omega_2)}\\
-Re{X^1(\omega_1)}].[Re{X^1(\omega_2)} -Re{X^0(\omega_2)}]
\end{aligned}
\end{equation}}

For further iteration, we have considered:
{\small
\begin{equation}\label{eq:31}
\begin{aligned}
\left[ {X}_{rec}^n(\omega)\right]_{1 \times (2N+1)}=\left[ {X}_{rec}^n(\omega_j)\right] \quad 1\leq j \leq 2N+1
\end{aligned}
\end{equation}}

Again, by expanding the signal with its complex conjugate:
{\small
\begin{equation}\label{eq:32}
\begin{aligned}
\left[ \hat{X}_{rec}^n(\omega)\right]_{1 \times (2N+1)}=\left[ {X}_{rec}^n(\omega_i), {X}_{rec}^n(\omega_{N-i}) \right] \quad 1\leq i \leq 2N+1
\end{aligned}
\end{equation}}

$\hat{X}_{rec}^n(\omega)$is the complex conjugate expand of ${X}_{rec}^n(\omega)$ that means the real part of ${X}_{rec}^n(\omega_i)$ is expanded even and the imaginary part of ${X}_{rec}^n(\omega_i)$ is expanded odd.

\begin{equation}\label{eq:33}
\begin{aligned}
\left[ \hat{X}_{rec}^n(\omega)\right]_{1 \times (2N+1)}=\left[ {X}_{rec}^n(\omega_j) \right] \quad 1\leq j \leq 2N+1
\end{aligned}
\end{equation}

\begin{equation}\label{eq:34}
\begin{aligned}
\left[ WFT \right]_{n_s \times (2N+1)}=\left[ \mu_k(s_k \omega_j)^m e^{s_k \omega_j} \right] \\ \quad 1\leq k \leq n_s \quad 1\leq j \leq 2N+1 \quad m=4
\end{aligned}
\end{equation}

Remark. $\mu_k$ is determined by the type of mother wavelet. For the $Paul4,  m=4$:
{\footnotesize
\begin{equation}\label{eq:35}
\begin{aligned}
{\left[ {{{{\mu }}_{{k}}}} \right]_{1 \times {{\rm{n}}_{\rm{s}}}}}{\rm{\;}} = \left| {\sqrt {{{\Delta \omega }}.\left( {2{\rm{N}} + 1} \right)} } \right|\sqrt {\frac{{{{\rm{s}}_{\rm{k}}}}}{{{\rm{m}}\left( {2{\rm{m}} - 1} \right)!}}} {{\;\;\;\;}},{{\;\;\Delta \omega }} = {{{\omega }}_{\rm{N}}} - {{{\omega }}_{{\rm{N}} - 1}}
\end{aligned}
\end{equation}}
	
Or we can rewrite the $WFT$ as follows:
{\tiny
\begin{flalign}\label{eq:36}
\begin{aligned}
\left[ WFT \right]_{n_s \times (2N+1)}=&\left[ WFT_{k,j} \right]\\
=
\begin{cases}
      \left [ \left|  \sqrt{\Delta \omega (2N+1)}\right| \sqrt{\frac{s_k}{m(2m-1)!}} (s_k \omega_j)^m e^{-s_k \omega_j} \right ] & \substack{1 \leq k \leq n_s, \quad 1\leq j \leq N \\ m=4} \\
      \\
      0 & \substack{1 \leq k \leq n_s \\ N+1\leq j \leq 2N+1}
   \end{cases}
\end{aligned}
\end{flalign}}


{\tiny
\begin{equation}\label{eq:37}
\begin{aligned}
\left[ {X}_{rec}^n(\omega)\right]_{n_s \times (2N+1)}=\left[ \left[ \hat{X}_{rec}^n(\omega)\right]_{1 \times (2N+1)} \right] \quad 1\leq j \leq 2N+1
\end{aligned}
\end{equation}
{\tiny
\begin{equation}\label{eq:38}
\begin{aligned}
\left[ W_r\right]_{n_s \times (2N+1)}=\left[ W_{r_{k,j}}\right] \\ = Re \left \{ IFFT \left \{ \left[ \left|  \sqrt{\Delta \omega (2N+1)}\right| \sqrt{\frac{s_k}{m(2m-1)!}} (s_k \omega_j)^m e^{-s_k \omega_j} \right] \times \hat{X}_{rec}^n(\omega_j)\right \} \right \}
\end{aligned}
\end{equation}}}

\begin{equation}\label{eq:39}
\begin{aligned}
\left[ s\right]_{n_s \times (2N+1)}=\left[ s_k \right]=\left[\left[\sqrt{s_k}\right]_{n_s \times 1}\right]
\end{aligned}
\end{equation}

\begin{equation}\label{eq:40}
\begin{aligned}
\left[P\right]_{n_s \times (2N+1)}=\frac{\left[W_r\right]_{n_s \times (2N+1)}}{\left[s\right]_{n_s \times (2N+1)}}=\frac{\left[W_{r_{k,j}}\right]}{\left[s_k\right]}
\end{aligned}
\end{equation}
{\tiny
\begin{equation}\label{eq:41}
\begin{aligned}
\left[ Sum\right]_{1 \times (2N+1)}=\left[ Sum_j\right] \\ = \mathlarger{\mathlarger{ \sum}}_{k=1}^{n_s} \frac{Re \left \{ IFFT \left \{ \left[ \left|  \sqrt{\Delta \omega (2N+1)}\right| \sqrt{\frac{s_k}{m(2m-1)!}} (s_k \omega_j)^m e^{-s_k \omega_j} \right] \times \hat{X}_{rec}^n(\omega_j)\right \} \right \}}{\sqrt{s_k}}
\end{aligned}
\end{equation}}
{\small
\begin{equation}\label{eq:42}
\begin{aligned}
\left[ {{{x}}_{{\rm{rec}}}^{{\rm{n}} + 1}\left( {\rm{t}} \right)} \right]_{1 \times 2{\rm{N}}} = \left[ {{\rm{x}}_{{\rm{rec}}}^{{\rm{n}} + 1}\left( {{{\rm{t}}_{\rm{j}}}} \right)} \right] = \frac{{\frac{{\left[ Sum_j \right]}}{C} - \tilde x_{rec}^{n + 1}}}{{{{{\mu }}_{{\rm{wav}}}}}} + \frac{{{X^0}\left( {{\omega _1}} \right)}}{N}
\end{aligned}
\end{equation}}

$\tilde{X}_{rec}^{n+1}$ is the mean value of the $\left[x_{rec}^{n+1} (t)\right]_{(1\times 2N)}$ vector. That means :

\begin{equation}\label{eq:43}
\begin{aligned}
\tilde{X}_{rec}^{n+1}=\frac{{X}_{rec}^{n+1}(t_1)+{X}_{rec}^{n+1}(t_2)+...+{X}_{rec}^{n+1}(t_{2N})}{2N}  \end{aligned}
\end{equation}

$\frac{X^0(\omega_1)}{N}$, according to the Final-value theorem is the mean value (DC value) of the time-domain signal of $x(t)$.

\begin{equation}\label{eq:44}
{\tiny
\begin{aligned}
\left[ x_{rec}^{n+1}(t)\right]_{1\times 2N}
= \frac{\mathlarger{\mathlarger{ \sum}}_{k=1}^{n_s} \frac{Re \left \{ IFFT \left \{ \left[ \left|  \sqrt{\Delta \omega (2N+1)}\right| \sqrt{\frac{s_k}{m(2m-1)!}} (s_k \omega_j)^m e^{-s_k \omega_j} \right] \times \hat{X}_{rec}^n(\omega_j)\right \} \right \}-\tilde{X}_{rec}^{n+1}}{C \sqrt{s_k}}}{\mu_{wav}} \\
+\frac{X^0(\omega_1)}{N}
\\
x_{rec}=\frac{1}{C\mu}\left(S-\hat{X}\right)+\alpha
\end{aligned}}
\end{equation}

\begin{equation}\label{eq:45}
\begin{aligned}
\left[{X}_{rec}^{n+1}(t)\right]_{1\times 2N}=\left[{X}_{rec}^{n+1}(t_1),...,{X}_{rec}^{n+1}(t_{2N})\right]
\end{aligned}
\end{equation}

To calculate $C_\psi$:

\begin{equation}\label{eq:46}
\begin{aligned}
{{{C}}_{{\psi }}} = {\rm{\;}}\int_{ - \infty }^{ + \infty } \frac{{{{\left| {{{\Psi }}\left( {{\omega }} \right)} \right|}^2}}}{{\left| {{\omega }} \right|}}{{d\omega }} < \infty
\end{aligned}
\end{equation}
{\tiny
\begin{equation}\label{eq:47}
\begin{aligned}
\left[ W\Delta \right]_{n_s \times 1}=\left[ W\Delta_{k} \right]=\frac{1}{2N+1} \mathlarger{\mathlarger{ \sum}}_{k=1}^{2N+1} WFT_{k,j}\\
= \frac{1}{2N+1} \mathlarger{\mathlarger{ \sum}}_{k=1}^{2N+1}
\begin{cases}
      \left [ \left|  \sqrt{\Delta \omega (2N+1)}\right| \sqrt{\frac{s_k}{m(2m-1)!}} (s_k \omega_j)^m e^{-s_k \omega_j} \right ] & \substack{1 \leq k \leq n_s, \quad 1\leq j \leq N \\ m=4} \\
      0 & \substack{1 \leq k \leq n_s \\ N+1\leq j \leq 2N+1}
   \end{cases}\\
\left[s_k \right]_{n_s \times 1}=\left[ \sqrt{s_k}\right]
\end{aligned}
\end{equation}}
{\tiny
\begin{equation}\label{eq:47}
\begin{aligned}
\left[ D \right]_{n_s \times 1}=\left[ D_{k} \right]=\frac{Re \left \{ \left[ W\Delta_k\right] \right \}}{\sqrt{s_k}} \\
= \frac{Re \left \{\frac{1}{2N+1} \mathlarger{\mathlarger{ \sum}}_{k=1}^{2N+1}
\begin{cases}
      \left [ \left|  \sqrt{\Delta \omega (2N+1)}\right| \sqrt{\frac{s_k}{m(2m-1)!}} (s_k \omega_j)^m e^{-s_k \omega_j} \right ] & \substack{1 \leq k \leq n_s, \quad 1\leq j \leq N \\ m=4} \\
      0 & \substack{1 \leq k \leq n_s \\ N+1\leq j \leq 2N+1}
   \end{cases} \right \}}{\left[ \sqrt{s_k}\right]}
\end{aligned}
\end{equation}}

{\tiny
\begin{equation}\label{eq:48}
\begin{aligned}
\left[ C \right]_{1 \times 1}=\mathlarger{\mathlarger{ \sum}}_{k=1}^{n_s} \left[ D_{k} \right] \\
= \mathlarger{\mathlarger{ \sum}}_{k=1}^{n_s} \frac{Re \left \{\frac{1}{2N+1} \mathlarger{\mathlarger{ \sum}}_{k=1}^{2N+1}
\begin{cases}
      \left [ \left|  \sqrt{\Delta \omega (2N+1)}\right| \sqrt{\frac{s_k}{m(2m-1)!}} (s_k \omega_j)^m e^{-s_k \omega_j} \right ] & \substack{1 \leq k \leq n_s, \quad 1\leq j \leq N \\ m=4} \\
      0 & \substack{1 \leq k \leq n_s \\ N+1\leq j \leq 2N+1}
   \end{cases} \right \}}{\left[ \sqrt{s_k}\right]}
\end{aligned}
\end{equation}}

\textbf{Remark:} To calculate  $\mu_{wav}$, we have used two tabular tables in MATLAB in CWTFT command developed by the Mathwork team using trial and error method and considering different types of signals. There are different tabular tables for the various wavelet types.

\begin{equation}\label{eq:49}
\begin{aligned}
\left[DD\right]_{1\times 37}=\left[DD_l\right]=\left[ tab_{val}-m \right], \quad  m=4\\
min= minimum(\left[DD\right]), \quad  index=l_{min}, \quad  \epsilon=2.22 \times 10^{-16}
\end{aligned}
\end{equation}
{\tiny
\begin{equation}\label{eq:50}
\begin{aligned}
\mu_{wav}= \\
\begin{cases}
      tab_{val}(index) & min <\sqrt{\epsilon} \\
      \frac{(tab{val}(index+1)-m)\times tab_{\mu}(index-1)+(m-tab{val}(index-1))\times tab_{\mu}(index+1)}{tab{val}(index+1)-tab{val}(index-1)} & min \geq \sqrt{\epsilon}
   \end{cases}
\end{aligned}
\end{equation}}
{\tiny
\begin{equation}\label{eq:51}
\begin{aligned}
\left[ x_{rec}^{n+1}(t)\right]_{1\times 2N} \\
= \frac{\mathlarger{\mathlarger{ \sum}}_{k=1}^{n_s} \frac{Re \left \{ IFFT \left \{ \left[ \left|  \sqrt{\Delta \omega (2N+1)}\right| \sqrt{\frac{s_k}{m(2m-1)!}} (s_k \omega_j)^m e^{-s_k \omega_j} \right] \times \hat{X}_{rec}^n(\omega_j)\right \} \right \}-\tilde{X}_{rec}^{n+1}}{C \sqrt{s_k}}}{\mu_{wav}} \\
+\frac{X^0(\omega_1)}{N}
\end{aligned}
\end{equation}}

\section{Error Analysis}

In this section, we study the convergence of the proposed algorithm and provide an error analysis.

\textbf{Theorem:}
\begin{equation}\label{eq:52}
\begin{aligned}
\lim_{n \to n_s}\left| x_{rec}^{n+1}(t)-x_{rec}^n(t) \right| \leq \epsilon
\end{aligned}
\end{equation}

\textbf{Proof:}
According to the described algorithm, we have:
\begin{equation}\label{eq:53}
\begin{aligned}
[X_{rec}^{n+1}(t)-X_{rec}^{n}(t)] \leq B \\
\end{aligned}
\end{equation}

where
{\tiny
\begin{equation}
B\le \frac{1}{\mu_{wav}}\mathlarger{\mathlarger{ \sum}}_{k=1}^{n_s} \frac{Re \left \{ IFFT \left \{ \left[ \left|  \sqrt{\Delta \omega (2N+1)}\right| \sqrt{\frac{s_k}{m(2m-1)!}} (s_k \omega_j)^m e^{-s_k \omega_j} \right] \times \hat{X}_{rec}^n(\omega_j)\right \} \right \}}{C \sqrt{s_k}}\\
 \end{equation}}

According to equation \eqref{eq:39}, the bound $B$ decreases as $n$ increases, owing to the exponential term in the bound. This proves the desired result.

\section{Approximation by Exponential Sums on Discrete and Continuous Domains}
\label{sec 9}

To generalize the application of the proposed method to the series of exponential functions, which appear in the major applications e.g. ship hull, we hire the following lemma. In this regard we consider:

\begin{equation}\label{eq:54}
\begin{aligned}
V_n (\mathbb{R})=\{ Y \in C^n(\mathbb{R}):[(D+\lambda_1)...(D+\lambda_n)]Y\}=0\\
for \quad some \quad  \lambda_1,...,\lambda_n \in \mathbb{R} \quad \quad
D=\frac{d}{dt}\nonumber
\end{aligned}
\end{equation}

\begin{equation}\label{eq:55}
\begin{aligned}
F=sup \left \{ \left| F(t)\right|: \quad 0\leq t \leq b \right \}
\end{aligned}
\end{equation}

\textbf{Lemma:}
Let $F \in \left[ 0,b \right]$ with $0<b<\infty$  and let $Y(t)=a_1 e^{\lambda_1 t}+ \cdot +a_k e^{\lambda_k t}$ where $\lambda_1 < \cdot <\lambda_k$  and  $a_i \neq 0$ for $i=1, \cdot , k$. Then $Y$  is a unique approximation to $F$ on $[0,b]$  from  $V_n(\mathbb{R})$ if and only if $F-Y$  alternates at least $n+k$  times on $[0,b]$ .

\textbf{Proof:} See \cite{daubechies1992ten}.

\section{Simulation Results}
Based on the discussion in \ref{sec 9}, for a signal that can be approximated by sum of exponential terms times a power series, we can apply this method:

\begin{equation}\label{eq:56}
\begin{aligned}
f(t)=\mathlarger{\mathlarger{\sum}}_{n=1}^{10} t^n e^{-n|t|}= te^{-|t|}+\frac{1}{2}t^2e^{-2|t|}+\cdot+\frac{1}{10}t^10e^{-10|t|}
\end{aligned}
\end{equation}

\begin{equation}\label{eq:57}
\begin{aligned}
coeff = Re \left \{X^n(\omega_0)\right \}+2 \log_{10} (\frac{\Delta \omega}{2 \pi})
ln (S_n)\times \\ sign [Re\left \{X^n(\omega_2)\right \}
-Re{X^n(\omega_1)}].[Re{X^{n+1}(\omega_2)} \\ -Re{X^n(\omega_2)}]
\end{aligned}
\end{equation}

\begin{figure}[!h]
\centering
\includegraphics[width=3in]{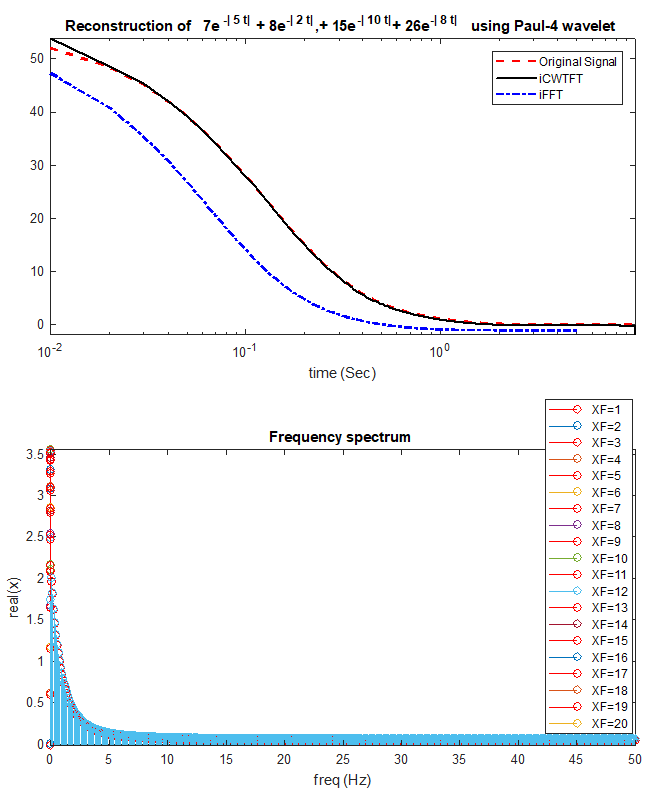}
 \caption{---}
\label{sim_res1}
\end{figure}

\begin{equation}\label{eq:58}
{\small
\begin{aligned}
coeff = Re \left \{X^n(\omega_0)\right \}+10 \log_{10} (\frac{\Delta \omega}{2 \pi})
ln (S_n)\times \\ sign [Re\left \{X^n(\omega_2)\right \}
-Re{X^n(\omega_1)}].[Re{X^{n+1}(\omega_2)} \\ -Re{X^n(\omega_2)}]
\end{aligned}}
\end{equation}

\begin{figure}[!h]
\centering
\includegraphics[width=3in]{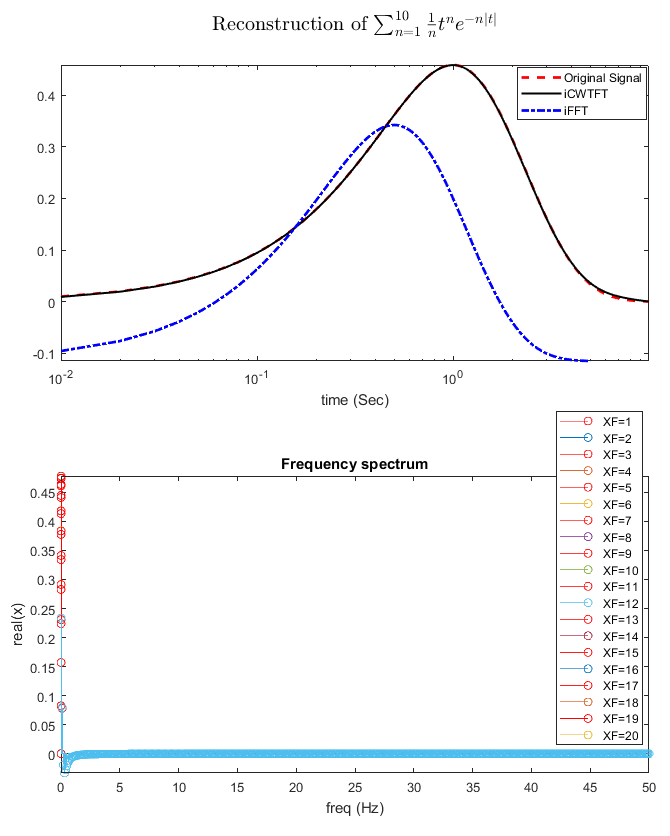}
 \caption{---}
\label{sim_res2}
\end{figure}

\begin{equation}\label{eq:59}
\begin{aligned}
coeff = Re \left \{X^n(\omega_0)\right \}+430 \log_{10} (\frac{\Delta \omega}{2 \pi})
ln (S_n)\times \\ sign [Re\left \{X^n(\omega_2)\right \}
-Re{X^n(\omega_1)}].[Re{X^{n+1}(\omega_2)} \\ -Re{X^n(\omega_2)}]
\end{aligned}
\end{equation}

\begin{figure}[!h]
\centering
\includegraphics[width=3in]{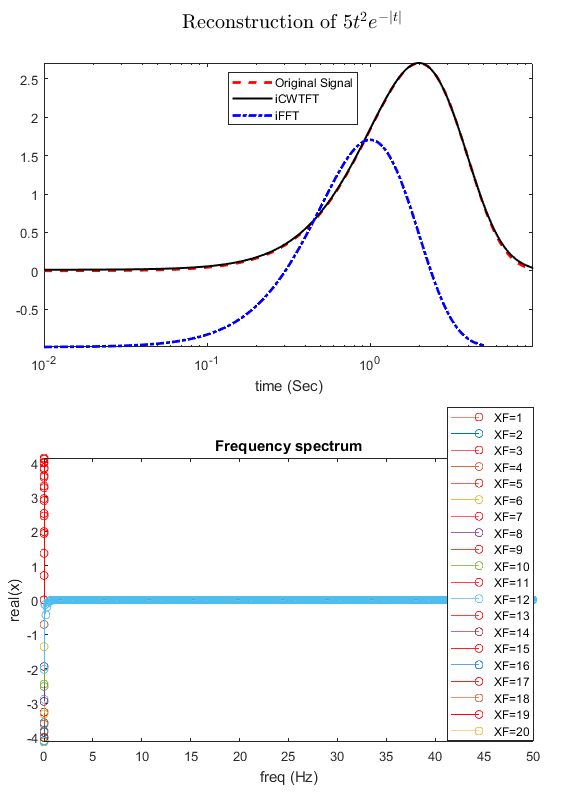}
 \caption{---}
\label{sim_res3}
\end{figure}

\section{Conclusion}
This paper presents a novel wavelet-based algorithm for reconstructing the time-domain impulse response from a band-limited set of Scattering parameters. The method is demonstrated using calculations of the US Naval electric-ship hull behavior, but since scattering parameters are commonly used to describe a system's behavior versus frequency and their related impulse responses in the time-domain, the proposed method can be used for a variety of systems.

The algorithm employs a recursive approach where the original data, in the form of the system's scattering parameter in the frequency-domain, is transformed using a special method of wavelet-Fourier transform. The wavelet transform of the frequency spectrum is calculated using the provided equations, which is equivalent to applying the wavelet transform to the related time-domain signal. The signal is then reconstructed in the time-domain using an inverse wavelet transform, and its Fourier transform is calculated to obtain the frequency spectrum. The missing DC frequency and other frequency components are retrieved using the provided coefficients, and this process is repeated for a determined number of iterations (with the related formula provided).

The error analysis section demonstrates that the iteration algorithm is converging and the error value is calculated analytically. A lemma is also presented to show that every signal can be approximated with a weighted sum of exponentials. In the result section, examples of these sums are depicted, and it is shown that the algorithm can successfully and accurately approximate the signal compared to the Fourier transform, assuming that the signal is missing the DC point.

This algorithm can be particularly useful for calculating the impulse response of corrupted spectra of systems, especially at the DC-side. Future work may involve the approximation of the impulse response from uneven frequency-sampled spectra.

\section*{Acknowledgment}
 The authors would like to thank the anonymous reviewers for their valuable comments and suggestions to improve the quality of the paper. They are also grateful to the Office of Naval Research (ONR) for their financial support of this work through grant
N00014-15-1-2276.

\bibliography{RYM10}
\bibliographystyle{IEEEtran}

%

\vskip 2pt plus -1fil

\begin{IEEEbiography}[{\includegraphics[width=1in,height=1.25in,clip,keepaspectratio]{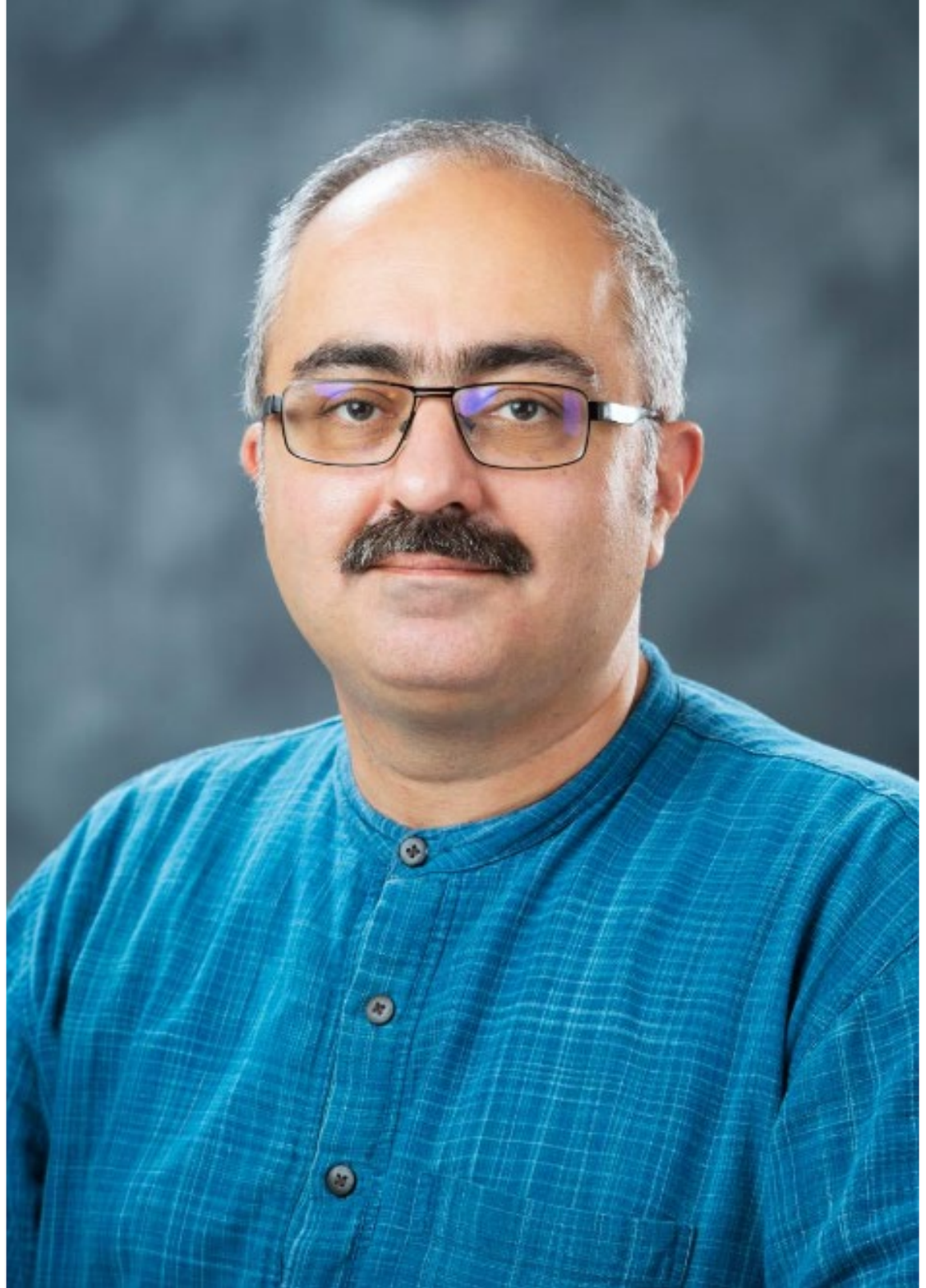}}]{Yarahmadian}
He earned his Ph.D. in Applied Mathematics from the Department of Mathematics at Indiana University, USA, in 2008. His research interests encompass partial differential equations, mathematical biology (with a focus on the dynamics of microtubules and Alzheimer's disease), stability of boundary layers, signal processing, and stochastic processes. He actively participates in interdisciplinary projects with the electrical engineering, mechanical engineering, civil engineering, computer science, and physics departments at Mississippi State University. In these collaborations, he applies a variety of analytical and computational techniques, including PDE (boundary layer theory), SDE, signal processing, random evolution, and non-parametric statistical methods. He is a member of the Society for Industrial and Applied Mathematics and has been serving as an associate professor at Mississippi State University since 2009.
\end{IEEEbiography}

\vskip 2pt plus -1fil

\begin{IEEEbiography}[{\includegraphics[width=1in,height=1.25in,clip,keepaspectratio]{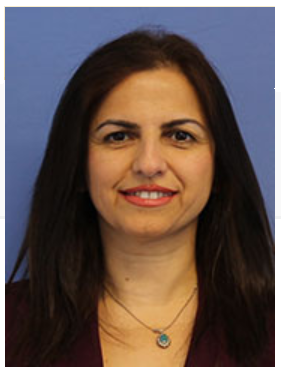}}]{Maryam Rahmani}
received her Ph.D. degree in Electrical Engineering from Mississippi State University, Starkville, MS, USA, in 2017. Her research interests include RF/Microwave circuit and module design, Antenna design, Electromagnetics, and RADAR systems. She served as a Visiting Assistant Professor in the Purdue University, Indianapolis, IN, USA. Since 2021, she has been a NASA Postdoctoral Fellow (NPP) at Goddard Space Flight Center (GSFC), conducting research in astrophysics on superconducting detectors and on-chip spectrometers.
\end{IEEEbiography}

\vskip -2\baselineskip plus -1fil

\begin{IEEEbiography}[{\includegraphics[width=1in,height=1.25in,clip,keepaspectratio]{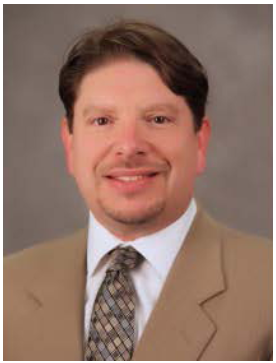}}]{Michael S. Mazzola}
(S’86–M’89) received the Ph.D. degree in electrical engineering from Old Dominion University, Norfolk, VA, USA, in 1990. He was an Endowed Professor of Electrical and Computer Engineering with Mississippi State University, Starkville, MS, USA, and an Associate
Director of the Center for Advanced Vehicular Systems (CAVS), Starkville, MS, USA until 2017. He was a cofounder of SemiSouth Laboratories, Inc., Starkville, MS, USA, a silicon carbide spin-out of Mississippi State University. Dr. Mazzola holds 14 patents.
His research interests include high-voltage engineering, power systems modeling and simulation, the application of silicon carbide semiconductor devices in power electronics and the control of hybrid electric vehicle power trains as well as energy, electricity, the environment and transportation. He has served as executive director of the Energy Production and Infrastructure Center (EPIC) at the University of North Carolina at Charlotte and the Duke Energy Distinguished Chair of Power Engineering Systems since 2017. Recently, Dr. Mazzola named as the executive director of the Future Use of Energy in Louisiana (FUEL), a statewide effort led by LSU to lead the clean energy transition and decarbonize the state’s industrial corridor.
\end{IEEEbiography}


\end{document}